\documentclass{PoS}

\usepackage{graphicx}
\usepackage{amsmath}
\usepackage{amssymb}
\usepackage{cite}

\title{Bounds on new physics from electric dipole moments}

\ShortTitle{Bounds on new physics from electric dipole moments}

\author{\speaker{Martin JUNG}\\%
        TUM Institute for Advanced Study / Excellence Cluster Universe,\newline Technische Universit\"at M\"unchen\\
        E-mail: \email{martin.jung@tum.de}\\
        Preprint: \email{FLAVOUR(267104)-ERC-109}
        
        }

\abstract{Electric dipole moments are extremely sensitive probes for additional sources of CP violation in new physics models. The multi-scale
 problem of relating the high-precision measurements with neutrons, atoms and molecules to fundamental parameters can be approached model-independently to a
large extent; however, care must be taken to include the uncertainties from especially nuclear and QCD calculations properly. The resulting bounds on
fundamental parameters are illustrated in the context of Two-Higgs-Doublet models.}

\FullConference{Flavor Physics \& CP Violation 2015\\
         May 25-29, 2015\\
         Nagoya, Japan}

\begin{document}

\section{Introduction}
Electric dipole moments (EDMs) provide a competitive means to search for new physics (NP), complementary to strategies like direct searches at hadron
colliders, but also to other indirect searches like the flavour-changing processes investigated at the flavour factories. The exceptional sensitivity
is due to the combination of experimental precision with a tiny Standard Model (SM) background. The smallness of the latter is related to the very
specific connection between flavour and CP violation in the SM,\footnote{EDMs are T,P-odd; their existence implies also CP violation when assuming CPT to be
conserved as we will in this article.} embodied by the Kobayashi-Maskawa mechanism~\cite{Kobayashi}, which is very effective in suppressing
flavour-changing neutral currents (FCNCs)~\cite{Glashow:1970gm}, and even more so in suppressing flavour-conserving ones involving CP violation. An
exception is provided by the gluonic operator $\mathcal{O}_{G\tilde G}\propto\epsilon_{\mu\nu\rho\sigma}G^{\mu\nu}G^{\rho\sigma}$, yielding a
potentially very large contribution to hadronic EDMs which is, however, strongly bounded experimentally, constituting the \emph{strong CP problem}.
To explain its absence, typically symmetries are invoked, involving additional particles. The most famous example is the \emph{Peccei-Quinn
mechanism}~\cite{Peccei:1977hh}, predicting the presence of axions~\cite{Weinberg:1977ma,Wilczek:1977pj}. While these have not yet been observed, it
is implicitly assumed in this work when discussing hadronic EDMs that the strong CP problem is solved by this or a similar mechanism.
The remaining SM contributions then lead to EDMs many orders of magnitude below the present
limits, \emph{e.g.} $d_n^{\rm SM,CKM}\lesssim 10^{-(31-32)}\,e\,{\rm
cm}$~\cite{Khriplovich:1981ca,Gavela:1981sk,McKellar:1987tf,Mannel:2012qk}.
Importantly, for leptonic EDMs no assumption regarding $\mathcal{O}_{G\tilde G}$ is necessary; the SM contribution to the electron EDM is estimated
to be  $d_e^{\rm SM}\lesssim10^{-38}\,e\,{\rm cm}$~\cite{Pospelov:1991zt,Booth:1993af,Pospelov:2013sca}.
The observation of an EDM with the present experimental precision would therefore clearly constitute a NP signal, especially in the leptonic sector.

Sakharov's conditions~\cite{Sakharov:1967dj} require the presence of new sources of CP violation to explain the observed baryon
asymmetry of the universe; while this does not \emph{imply} sizable EDMs, they are generally very sensitive to such sources. 
In fact, generic NP scenarios usually yield contributions that are large compared to experimental limits, implying either a high NP scale or a very
specific structure for additional CP-violating contributions, similar to the situation in the flavour-changing sector. Casting these
qualitative statements into reliable bounds on model parameters requires knowledge of their relation to the experimental observables --
typically (bounds on) frequency shifts obtained for composite systems. The calculation of these relations proceeds via a series of effective field
theories (EFTs), see \emph{e.g.} Refs.~\cite{Pospelov:2005pr,Engel:2013lsa} for recent reviews and references therein. Importantly, this approach
allows to perform a large part of the analysis model-independently. The calculation of the matrix elements of the corresponding effective operators
often include large uncertainties which have to be taken into account, see Refs.~\cite{Jung:2013hka,Engel:2013lsa} for recent detailed discussions.
Furthermore, in composite systems different contributions can exhibit cancellations; this issue can already be systematically addressed for
paramagnetic systems~\cite{PhysRevA.85.029901,Jung:2013mg}, and in the future potentially also for diamagnetic ones~\cite{Chupp:2014gka}.

This article proceeds as follows: model-independent constraints are discussed in the following section, focusing on paramagnetic systems. In
Sec.~\ref{sec::EDMsandNP} NP contributions to EDMs are discussed first generally, then using Two-Higgs-Doublet models (2HDMs) as a specific example.
We conclude in Sec.~\ref{sec::Conclusions}.

\section{Model-independent constraints from EDM measurements}
The available competitive observables, that is, the EDMs of thorium monoxide (ThO) and ytterbium fluoride (YbF)
molecules~\cite{Baron:2013eja,Hudson:2011zz}, thallium (Tl) and mercury (Hg) atoms~\cite{Regan:2002ta,Griffith:2009zz} and the
neutron~\cite{Baker:2006ts,Afach:2015sja} (see also \cite{Serebrov:2013tba}), are related by calculations on the molecular, atomic, nuclear and QCD
levels to the coefficients of an EFT at a hadronic scale (see, \emph{e.g.}, Refs.~\cite{Khriplovich:1997ga,Pospelov:2005pr,Engel:2013lsa}):
\begin{eqnarray}\label{eq::Leff}
\mathcal{L}^{\rm EDM}_{\rm eff} = -\!\!\sum_{f}\frac{d_f^\gamma}{2}\mathcal{O}_f^\gamma-\sum_q\frac{d_q^C}{2}\mathcal{O}_q^C+C_W
\mathcal{O}_W+\sum_{f,f'}C_{ff'}\mathcal{O}_{ff'}\,.
\end{eqnarray}
This operator basis consists of (colour-)EDM operators $\mathcal{O}^{\gamma,C}_f$, the purely gluonic Weinberg operator $\mathcal{O}_W$ and T-
and P-violating four-fermion operators $\mathcal{O}_{ff'}$ without derivatives ($f^{(\prime)}=e,q$, $q=u,d,s$). Since these calculations do not depend
on the NP model under consideration, this Lagrangian is used as the interface between experiment and high-energy calculations: the latter
provide the model-specific expressions for the Wilson coefficients in Eq.~\eqref{eq::Leff}, with at least one more intermediate EFT at the electroweak
scale.

\begin{figure}
\includegraphics[width=6.2cm]{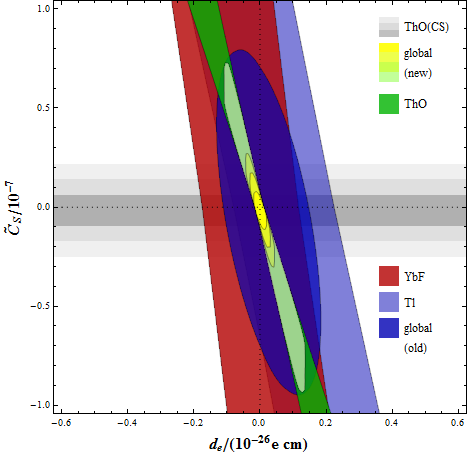}\qquad\qquad\qquad\includegraphics[width=6.2cm]{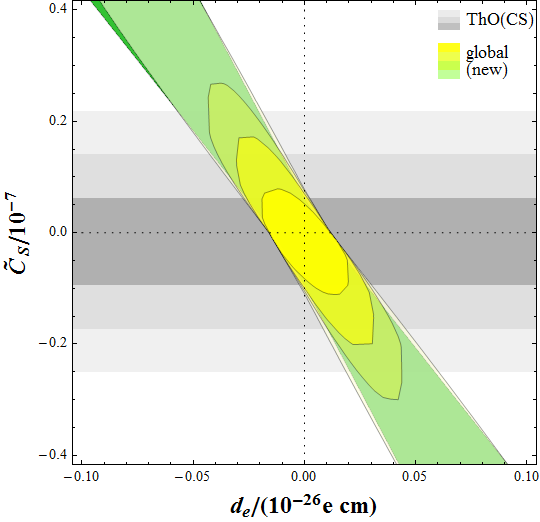}
\caption{\label{fig::globalFit} Fit to the recent measurements for paramagnetic
systems~\cite{Baron:2013eja,Hudson:2011zz,Regan:2002ta}, using additionally the Hg measurement~\cite{Griffith:2009zz} (blue ellipse on the left) or
the assumption of limited cancellations in ThO with $n=1,2,3$ (grey bands, both plots), see text. Plots taken from
Ref.~\cite{Jung:2013hka}.
}
\end{figure}

In neutral composite systems, the EDMs of the components are shielded; for non-relativistic, point-like constituents this shielding is perfect,
therefore measurements for this type of system rely on the violation of these assumptions~\cite{Schiff:1963zz}. For paramagnetic systems, relativistic
effects can actually lead to \emph{enhancement} factors, if the proton number $Z$ is large
enough~\cite{Sandars:1965xx,Sandars:1966xx,Flambaum:1976vg}, since two contributions scale approximately with $Z^3$: these are the ones from the
electron EDM and the scalar electron-nucleon coupling, $\tilde C_S$.\footnote{Note that $\tilde C_S$ depends in general on the considered system.
However, for the systems at hand (and more generally for heavy paramagnetic systems), it is universal to very good approximation~\cite{Jung:2013mg}.}
Heavy paramagnetic systems can therefore be assumed to be completely dominated by these two contributions, allowing a model-independent fit to
bound and eventually determine \emph{both} contributions, without the assumption of a vanishing electron-nucleon
contribution~\cite{PhysRevA.85.029901,Jung:2013mg}. In practice, there are two complications with this approach at present, which can however be
overcome with additional measurements. Firstly, the ratio of the coefficients of the two contributions is necessarily similar for heavy paramagnetic
systems~\cite{PhysRevA.85.029901}. This problem can be solved by performing measurements with atoms or molecules with largely different proton
numbers, such as rubidium and francium atoms. Absent such (competitive) measurements, it is possible to assume \emph{e.g.} the limit from Hg to
be saturated by the $\tilde C_S$ contribution~\cite{Jung:2013mg}: this is a conservative procedure, since the EDM of this system is
typically dominated by colour-EDM (cEDM) contributions. While this lead in the past to a similar limit on $d_e$ like  naively setting $\tilde
C_S\to 0$, the unmatched precision of the recent ThO measurement (in the $d_e$-$\tilde C_S$ plane) poses the second problem, namely that the
projections of the 2-dimensional fit remain almost unchanged, see Fig.~\ref{fig::globalFit}~(left). Since this may appear overly conservative, a
temporary option is to limit the cancellation between the two contributions within the ThO measurement, \emph{i.e.} to allow the $\tilde C_S$
contribution to at most saturate the experimental limit $n=1,2,3,\ldots$ times, see Fig.~\ref{fig::globalFit}~(right), as a compromise between the
conservative procedure described previously and the potentially critical assumption of $\tilde C_S=0$. The corresponding fits lead to 
\begin{equation}
d_e\leq1.0(0.16)\,10^{-27}e\,{\rm{cm}}\,\,(95\%~{\rm CL})\,,
\end{equation} 
using Hg and $n=1$, respectively. These values should be used when extracting bounds on
parameters from $d_e$ in any model in which the electron-nucleon contribution cannot be argued to be negligible. 
Furthermore, these 2-dimensional constraints allow to obtain
model-independent constraints on the EDMs of all other heavy paramagnetic systems~\cite{Jung:2013mg}; examples are 
\begin{equation}
d_{Rb}\leq 3(0.5)\times 10^{-26}e\,{\rm cm}\,,\quad d_{Cs}\leq 10(2)\times 10^{-26}e\,{\rm{cm}}\,,\quad  d_{Fr}\leq
60(20)\times 10^{-26}e\,{\rm cm}\,\,({\rm all}~95\%~{\rm CL}).
\end{equation}
A significant measurement in one of these systems larger than the first bound would indicate an experimental problem, while a value larger than the
bound in brackets would indicate large cancellations between the two contributions. These limits are orders of magnitude below existing experimental
ones~\cite{PhysRev.153.36,RbEDMthesis,Murthy:1989zz}. Importantly, present experiments aim at an even better
sensitivity~\cite{Weissetal,Amini:2007ku,2004APS..DMP.P1056K,Sakemi:2011zz,PhysRevX.2.041009}.

The extension of this type of fit to all EDM measurements is clearly possible and has been proposed in Ref.~\cite{Chupp:2014gka}. While this is
complicated by the many potential contributions -- all of the terms in Eq.~\eqref{eq::Leff} are relevant in general --, it should be aimed at in the
future. Since model-independent bounds/determinations are necessary to determine the specific structure of CP-violating NP contributions -- and
thereby potentially the model itself --, many measurements in different systems are very important. An additional complication for the EDMs of
diamagnetic systems and neutrons is that the theoretical uncertainties for the relevant matrix elements are often large and can in some cases preclude the extraction of conservative limits, for
instance on the cEDMs in Eq.~\eqref{eq::Leff} from Hg~\cite{Jung:2013hka}, further motivating complementary measurements.

\section{\label{sec::EDMsandNP} NP contributions to EDMs}
Reliable limits on parameters in NP models are difficult to achieve. Reasons are, apart from the fact discussed previously that presently less
competetive measurements than relevant effective operators exist, the presence of several contributions to each of these coefficients and the
various relevant hierarchies, \emph{i.e.} mass scales, small mixing parameters and loop factors. This complicates semi-model-independent analyses for
classes of models and allows strict statements only under additional assumptions. 

Generic NP contributions at tree- and one-loop level are in conflict with the stringent experimental limits. On the two-loop level, usually so-called
Barr-Zee- and Weinberg diagrams dominate~\cite{Weinberg:1989dx,Dicus:1989va,Barr:1990vd,Gunion:1990iv}, which compensate the additional loop factor by
avoiding small mass factors. However, flavour sectors are usually far from ``generic''; therefore in some cases also tree-level diagrams can be
relevant, for example when they involve small mass factors, see below.

In order to demonstrate these qualitative statements in a specific model, we consider
a general 2HDM, whose charged current Yukawa couplings can be parametrized in the Higgs basis as follows:
\begin{equation}\label{eq::LHcharged}
\mathcal L_Y^{H^\pm} \! =\! - \frac{\sqrt{2}}{v}\, H^+ \left\{  \bar{u}   \left[  V \varsigma_d M_d \mathcal P_R - \varsigma_u\, M_u^\dagger V \mathcal P_L  \right]   d\, +\,  \bar{\nu} \varsigma_l M_l \mathcal P_R l\right\} + \;\mathrm{h.c.} \, ,
\end{equation}
where the $M_i$ are diagonal mass matrices, $V$ denotes the CKM matrix, and the $\varsigma_f$ in principle arbitrary complex matrices. Below the
constraints are given for the elements of these matrices, which can be translated into the parameters of any given 2HDM. To be specific and
able to relate the resulting bounds also to those from other observables, we will furthermore consider the Aligned 2HDM
(A2HDM)~\cite{Pich:2009sp,Jung:2010ik}, where the $\varsigma_i$ are complex numbers, thereby avoiding FCNCs on tree level while still allowing for a
rich phenomenology including additional CP-violating phases.
The couplings of the neutral Higgs mass-eigenstates  $\varphi_i^0=\{h,H,A\}$ to a fermion species $f$ are denoted by $y_f^{\varphi_i}$; they depend
not only again on the $\varsigma_i$, but in general additionally on the parameters of the Higgs potential, severely complicating the analysis. These
couplings fulfill the relations
\begin{equation}\label{eq::ycancellation}
\sum_i {\rm Re}\left(y_f^{\varphi^0_i}\right){\rm Im}\left(y_{f'}^{\varphi^0_i}\right)=\pm{\rm Im}\left[(\varsigma_{F(f)}^*)_{ff}(\varsigma_{F(f')})_{f'f'}\right]\,,
\end{equation}
with a vanishing right-hand side for real $\varsigma_i$ (as \emph{e.g.} the case for $\mathcal Z_2$ 2HDM models) or $f=f'$ (more generally fermions of
the same family if the $\varsigma_i$ are family-universal as \emph{e.g.} in the A2HDM). Importantly, the right-hand side is independent of the
parameters of the scalar potential. While in practical calculations there are mass-dependent weight factors in the sum on the left, the relation still
holds exactly in two limits: trivially when the neutral scalars are degenerate, but also in the decoupling limit~\cite{Jung:2013hka}. Therefore, in
general cancellations can be expected for any mass spectrum and the influence of CP violation in the potential is reduced. Clearly, this observation
provides a protection against large EDMs for models which exhibit new CP-violating parameters in the potential, only. Below we will assume this
cancellation to occur and evaluate the right-hand side with a common weight factor at an intermediate effective neutral Higgs mass
$\overline{M}_\varphi$.

In this setup, the situation is typically the one described above: four-quark (tree-level) contributions are subleading, one-loop contributions to
(colour-)EDMs are under control  (but not necessarily tiny), and two-loop contributions are dominant, but also
the tree-level quark-electron couplings are relevant, despite the small mass factors~\cite{Buras:2010zm,Jung:2013hka}.

The electron EDM receives contributions mostly from Barr-Zee diagrams. The resulting constraints on ${\rm
Im}(\varsigma_{u,33}\varsigma_{l,11}^*)$ demonstrate the strength of this observable: ${\rm
Im}(\varsigma^*_{u,33}\varsigma_{l,11})\lesssim0.05$ (for $n=1$). This already questions the common assumptions that these factors are $\mathcal
O(1)$.
In the A2HDM this can be compared to the absolute value of this parameter combination obtained from leptonic and semileptonic
decays~\cite{Jung:2010ik,Celis:2012dk}, which is about a factor 1000 weaker.

As mentioned above, also the constraint from $\tilde C_S$ is relevant: while in this case the constraint seems much weaker, ${\rm
Im}(\varsigma^*_{d,11}\varsigma_{l,11})\lesssim15$, the bound on the ratio with $\bar M_\varphi^2$ is again at least a factor 100 stronger than the
corresponding one with the charged-Higgs mass from (semi-)leptonic processes~\cite{Jung:2010ik,Celis:2012dk}.

For the neutron, the constraint induced in the charged-Higgs sector via the Weinberg operator is the dominant one, leading to ${\rm
Im}(\varsigma^*_{u,33}\varsigma_{l,33})\lesssim1$. While this does not imply any sizable finetuning, it already prohibits large CP-violating
effects in other observables in specific models. For instance, while the indirect constraint from the branching ratio in $b\to s\gamma$ in the A2HDM
still allows for a sizable CP asymmetry for this process, a NP contribution of $|A_{\rm CP}(b\to s\gamma)|\lesssim1\%$ follows from the EDM bound
and the discussion in Refs.~\cite{Jung:2010ab,Jung:2012vu}.

These examples show the potential of EDMs, but also their complementarity to other searches, since only the imaginary parts of parameter combinations
are constrained. However, for the combinations EDMs are sensitive to, they are often the strongen constraints available.

\section{\label{sec::Conclusions} Conclusions}
EDMs provide unique constraints for the CP-violating sectors of NP models. A potential discovery of a non-vanishing fundamental EDM in any system
would be a major achievement, independent of its source. The interpretation of bounds and potential measurements in terms of fundamental theory
parameters requires the careful estimation of theoretical uncertainties and is complicated by potential cancellations on various levels. While this
problem can be addressed for heavy paramagnetic systems to extract the electron EDM and scalar electron-nucleon coupling model-independently, a
similar approach including all relevant systems should be aimed at, but requires several additional measurements for different systems.

For the occuring combinations of parameters, EDMs typically provide the most stringent constraints. We demonstrated this explicitly for
general 2HDMs, and more specifically for the A2HDM, where large CP-violating effects in other observables are strongly bounded by the existing EDM
limits. Given the present strength of the constraints, forthcoming experiments will test a crucial part of the parameter space and might turn existing
bounds into observations.

\bibliography{EDM_proceedings2}

\end{document}